\documentclass[sigconf]{acmart}
\acmConference[SPLC'19]{23rd International Systems and Software Product Line Conference}{9--13 September, 2019}{Paris, France}

\AtBeginDocument{%
  \providecommand\BibTeX{{%
    \normalfont B\kern-0.5em{\scshape i\kern-0.25em b}\kern-0.8em\TeX}}}

\copyrightyear{2019} 
\acmYear{2019} 
\acmDOI{10.1145/3307630.3342387}

\usepackage{listings}
\usepackage{subcaption}

\begin{document}

%
\title[Understanding Conditional Compilation]{Understanding Conditional Compilation Through Integrated Representation of Variability and Source Code}

%
\author{David Baum}
\email{david.baum@uni-leipzig.de}
\author{Christina Sixtus}
\email{wir13dvx@studserv.uni-leipzig.de}
\author{Lisa Vogelsberg}
\email{ges11eso@studserv.uni-leipzig.de}
\author{Ulrich Eisenecker}
\email{eisenecker@wifa.uni-leipzig.de}
\affiliation{%
  \institution{Leipzig University}
  \city{Leipzig}
  \country{Germany}
}

%

%
\begin{abstract}
The C preprocessor (CPP) is a standard tool for introducing variability into source programs 
and is often applied either implicitly or explicitly for implementing a Software Product Line (SPL).
Despite its practical relevance, CPP has many drawbacks.
Because of that it is very difficult to understand the variability implemented using CPP.
To facilitate this task we provide an innovative analytics tool which bridges the gap between feature models as more abstract representations of variability and its concrete implementation with the means of CPP.
It allows to interactively explore the entities of a source program with respect to the variability realized by conditional compilation.
Thus, it simplifies tracing and understanding the effect of enabling or disabling feature flags.
\end{abstract}


%
%


%

%

%
\maketitle

\section{Introduction}

Conditional compilation is a way of introducing variability to C source code immediately before compile time. 
The CPP can be used to include or exclude source code components, which change the structure and behavior of the resulting program.
Often Boolean feature flags are used to design complete SPLs.
The complexity created by the numerous variants is challenging.
Although feature models help to describe the variability, they are of limited use when working with the source code directly, e.g., during bug fixing.
In general, bugs that lead to unwanted runtime behavior are often more difficult to detect and to fix than compile time errors.
This applies even more if a bug only occurs under certain feature configurations.
For this reason, the developer needs support for answering the following questions, that appear regularly during development:
\textbf{Q1:} What effect does the activation of a feature have on the structure of a program?
\textbf{Q2:} Which elements are contained in the source code given a certain feature configuration?
%
With these questions in mind we developed an interactive analytics tool that provides the following functionality:
\begin{enumerate}
 \item It provides an overview over the structure of the system, i.e., all functions, global variables, and complex types that can be part of any variant.
 
 \item The user can define a set of flags and explore the structure of the resulting variant. 
        This includes method calls, read and write operations, as well as the original C code. 

 \item The analysis runs fully automated without any manual preparation steps.
\end{enumerate}

A demo is available online\footnote{\url{http://softvis.wifa.uni-leipzig.de/splc2019}}.
Additionally, the usage of the tool is demonstrated in a screencast\footnote{\url{http://softvis.wifa.uni-leipzig.de/splc2019screencast}}. 
We first examine how existing tools support the presented use case, followed by a presentation of our tool.
We will address several design choices, including variability extraction, and the visualizations the user interface is based on.
A small application scenario based on the online demo is presented in chapter 5.
Finally, we will discuss our previous experiences with the tool and future development.

\section {Related Work}



Most work on SPLs and variability either focuses on automatic checks at compile time or provides abstract models without a direct connection to the source code.
In the area of C code refactoring numerous works can be found, that take preprocessor statements into account~\cite{Garrido.2005,Spinellis.2010,M.Vittek.2003,Waddington.2005,Baxter.2001}.
Feature models are often used to prepare the extracted information.
Badros and Notkin have written a tool that analyzes unpreprocessed C source code with simple scripts~\cite{Badros.2000}.
The SPL community offers a number of tools for visualization and for better understanding variability points and variants.
For example, two Eclipse plugins visualize feature models and perform type checking of preprocessor code~\cite{Medeiros.2013, Thum.2014}.
With the help of Meta Programming System (MPS) different views for editing and understanding SPL source code can be provided to developers~\cite{Behringer.2017}.
Other tools generally support the development of SPLs without the need for specific focus on C source code.
Feigenspan et al. have developed an Eclipse plugin that enables highlighting of feature code~\cite{Feigenspan.2010}.
Nestor et al. have created visualizations for the configuration of SPLs~\cite{Nestor.2008}, but they do not provide a direct connection to the source code. 
The Feature Relation Graph presents possible feature combinations depending on a selected feature~\cite{Martinez.2014}. 
Illescas et al. as well as Urli et al. show different visualization models for feature combinations but without a connection to the source code~\cite{Illescas.2016, Urli.2015}.
Many works are based on the same extraction tools such as FeatureCoPP~\cite{Kruger2016}, SuperC~\cite{Gazzillo2012}, TypeChef~\cite{Kastner2011}, and Yacfe~\cite{Padioleau}. 
We are not aware of any tool that supports the presented use case satisfactorily.
In the area of SPLs, the focus of research is on the representation of variability points and legal combinations of features. 
In most cases links to the underlying source code are not presented.

In contrast, some tools are aimed at improving the developer's understanding of the code.
Livadas and Small have created an integrated development environment (IDE) extension that can be accessed by clicking a macro expansion.
It shows where a macro has been defined and how the macro is expanded~\cite{Livadas.1994}. 
Also Kullbach and Riediger visualize macro expansion and conditional compilation with an IDE extension by so-called folding. 
When clicking on a preprocessor instruction, the corresponding precompiled source code is collapsed~\cite{Kullbach.2001}.
However, these tools are only useful for local contexts and do not address systemwide variability.





\section{Variablity Extraction}

Comprehensive preprocessing of the C code and the CPP statements is required to provide answers to the questions that have been raised.
Our requirements on such a parser can be summarized as follows: 
\begin{enumerate}
 \item 
    The result of the parsing must contain all the linguistic means of the C standard. 
    This includes translation units, functions, elementary types, complex types, information about function calls as well as reading and writing of global variables.
\item
    The parser should consider the included files to handle declarations correctly.
\item
    Macro expansions should be performed before parsing since the content of the macros may influence feature detection and location.
\item
    The result of the parsing should contain information about the conditional compilation, including the CPP directives extracted from the source code. 
    Even more useful would be an evaluation of nested conditions and an explicit representation of alternatives as distinct branches in the result.
\end{enumerate}

There exist various tools with different scopes to analyze code with conditional compilation. 
We came to the conclusion that TypeChef meets our requirements best, although it is significantly slower than, e.g., SuperC. 
The goal of the developers of TypeChef was to create a complete and solid parser that can parse C code without manual preprocessing.
It uses an LL parser to create an abstract syntax tree (AST) which contains all of the variability information we need. 
We modified TypeChef to serialize the complete AST to an XML file for further processing with jQAssistant. 
This is a program for analyzing and visualisizing software artifacts~\cite{Muller2018}.
It is built on top of Neo4j, a graph database. 
%
We implemented a plugin for TypeChef to include C code and feature flags.
The result is a graph containing all code entities, method calls, read and write accesses, features, and their dependencies.

\section{User Interface}

Getaviz~\footnote{https://github.com/softvis-research/Getaviz} is an open source toolkit for visual software analytics~\cite{Baum2017}.
It uses jQAssistant as information source and supports the automatic generation of visualizations for different use cases~\cite{Baum2018}.
Getaviz comes with a highly configurable browser-based user interface for viewing and interacting with a visualization. 
Getaviz can be easily expanded to support new visualization types and interaction components. 
Hence, we used Getaviz as starting point and customized it to fit our requirements.

\begin{figure*}[h]
  \centering
  \includegraphics[width=0.95\linewidth]{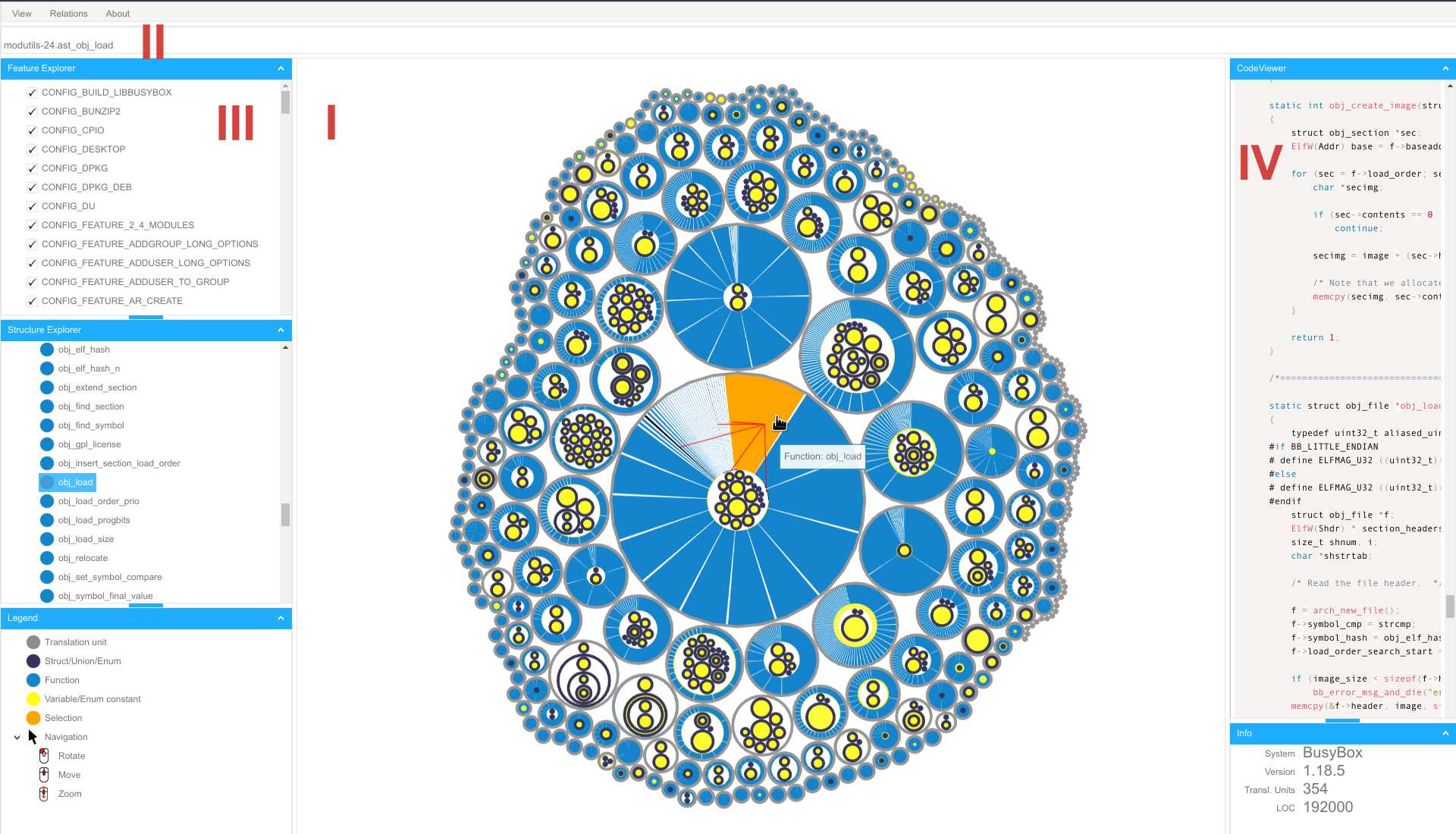}
\caption {Screenshot of Getaviz visualisizing the structure of BusyBox}
  \label{fig:screenshot}
\end{figure*}

Figure~\ref{fig:screenshot} shows the default view containing a visualization of the structure (I), a search bar (II), the \textit{FeatureExplorer} (III), and the \textit{CodeViewer} (IV).
To understand the structure and the included variablity it is useful to get an overview of the complete system first.
Therefore, we visualize the structure in such a way that it can be fully grasped at a glance.
This view contains all code entities that could be potentially compiled. 
Our prototype is based on the Recursive Disk (RD) metaphor~\cite{Mueller2015b}.
It is designed to visualize the structure of imperative programming languages, with an emphasis on object-oriented languages, especially Java.
As the name indicates, an RD visualization consists of nested disks, where each disk represents a package or a class in Java. 
In order to apply the visualization to C code, we had to make several changes.
We chose translation units as top level elements replacing packages. 
They are depicted as gray disks as shown in Figure~\ref{fig:conditional_compilation}.
A translation unit can contain multiple structs, unions, enums, global variables, and functions. 
Functions are depicted as blue segments. 
The area of a blue segment is proportional to the lines of code of the corresponding function. 
Variables are depicted as yellow segments that have a fixed size. 
Structs, enums, and unions are depicted as purple disks.
They can contain further elements according to the content of the C entities.
We have retained the original layout algorithm.
All disks are ordered by size and then placed spiral-shaped clockwise around the largest disk.
Although at first glance it seems chaotic, the emerging visual patterns and empty spaces give each disk a unique appearance and help the user to recognize specific disks.

The visualization is interactive, so the developer can easily explore it. 
The \textit{FeatureExplorer} contains all extracted feature flags of the system.
The developer can select or deselect individual flags and the visualization gets updated accordingly.
If the code entity is to be excluded by the CPP, then the graphical representation will be displayed transparently.
In this way, the user can explore and understand the impact of the different flags to answer Q2 without having to jump from source file to source file and manually evaluate macros.

Detail information is provided as tooltip.
In Figure~\ref{fig:screenshot}, the method \lstinline!obj_load! is selected and therefore highlighted orange.
The red lines represent method calls and variable accesses of this method.
Nevertheless, the source code is still of great interest for the developer since it is the main artifact to work with. 
To provide more context it is possible to view the source code directly in Getaviz.
The \textit{CodeViewer} on the right side displays the source code of the selected entity.

\section{Application}

To demonstrate the usefulness of our tool we chose BusyBox 1.18.5 since it is a highly customizable system.
It contains several hundreds of explicitly declared Boolean compile-time configuration options with complex dependencies~\cite{Kastner2012}.
One of these feature flags is \lstinline!CONFIG_DESKTOP! which affects the macros \lstinline!ENABLE_DESKTOP!, \lstinline!IF_DESKTOP!, and \lstinline!IF_NOT_DESKTOP!.
They are used in $75$ out of $354$ translation units.
Therefore, enabling this feature flag potentially changes behavior of more than $20\%$ of the system in a variety of locations that can not be easily traced by the developer.
Our tool takes the affected macros into account automatically and visualizes the influence of the feature flag on the structure with just one click. 

\begin{figure*}[h]
\centering
\begin{subfigure}[t]{0.3\textwidth} 
    \includegraphics[width=\textwidth]{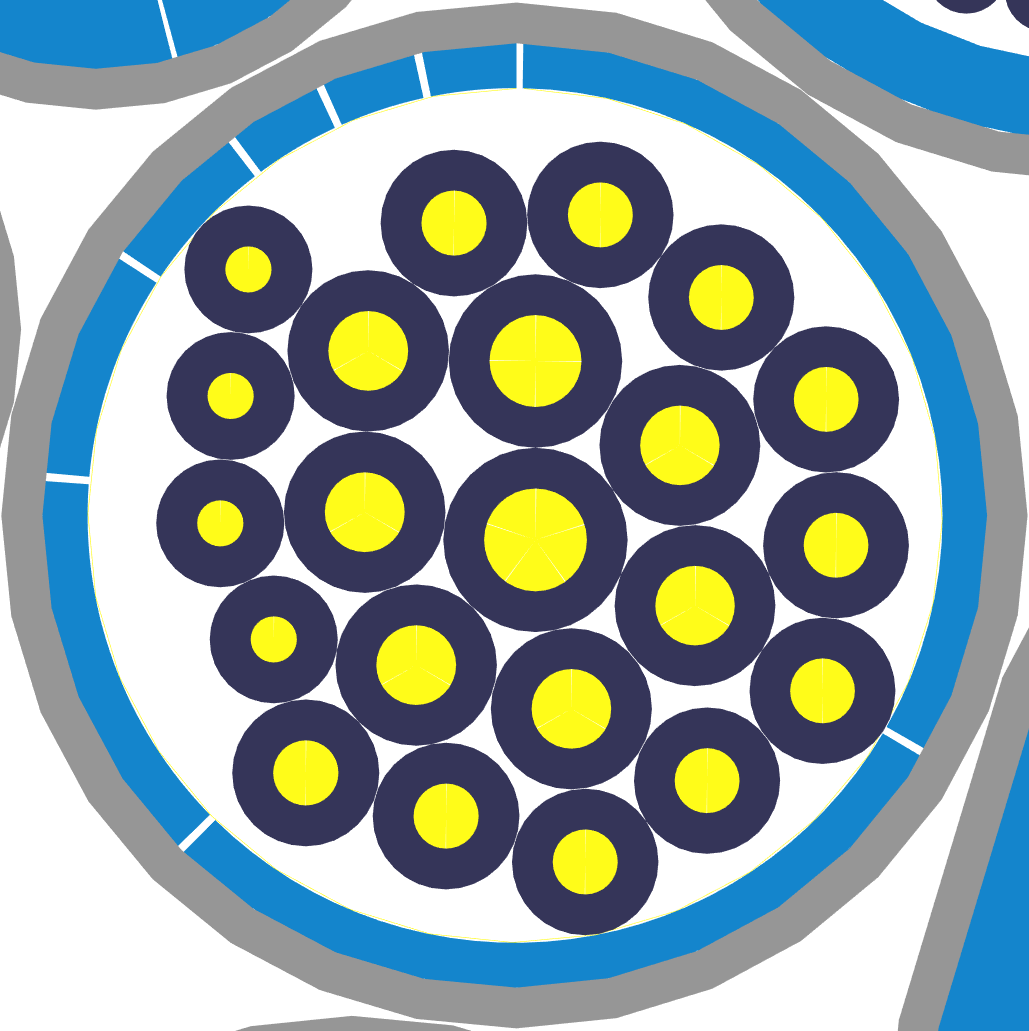}
    \subcaption { All \lstinline!find! specific features are enabled }
    \label{fig:screenshot_full}
\end{subfigure}
\hfill
\begin{subfigure}[t]{0.3\textwidth}
    \includegraphics[width=\textwidth]{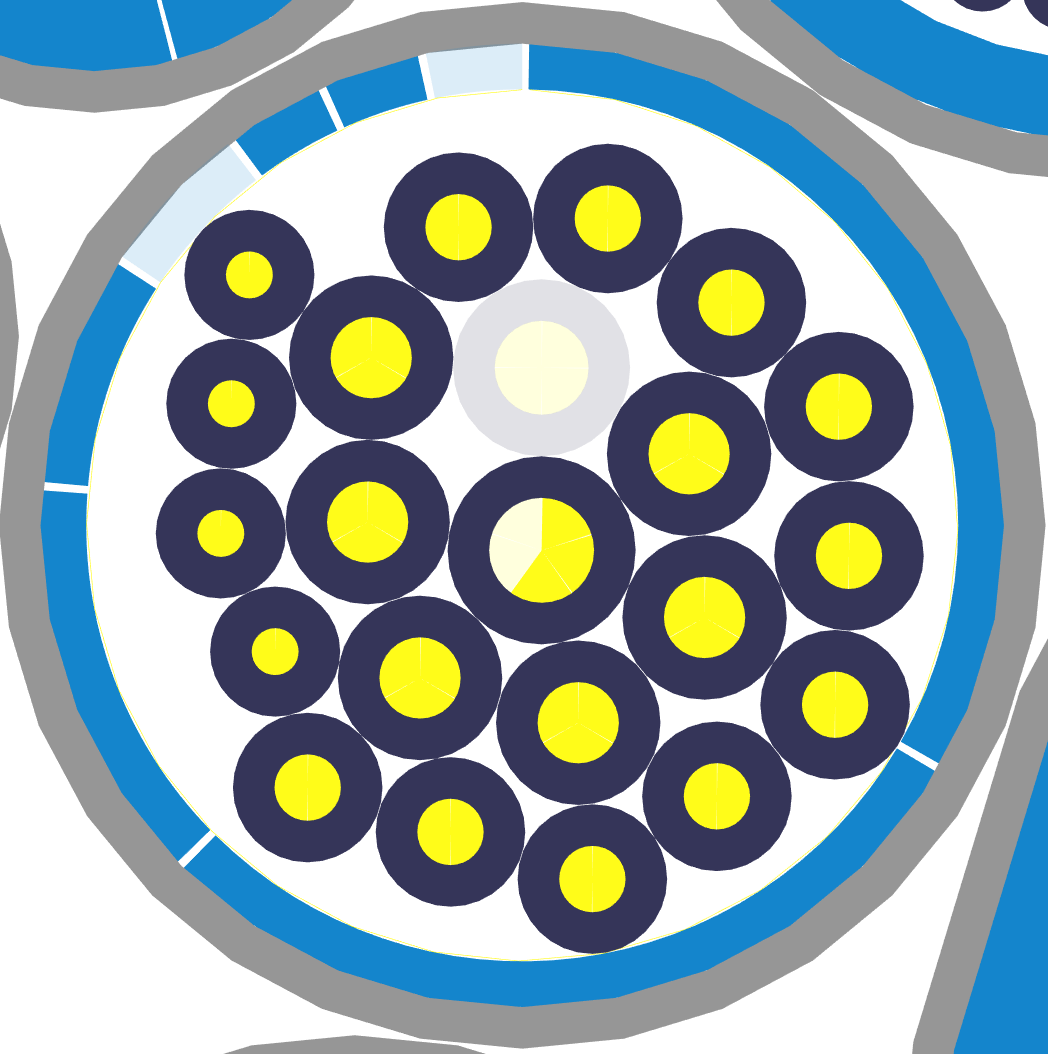}
    \subcaption { All \lstinline!find! specific features are enabled except \lstinline!FEATURE_FIND_EXEC! and \lstinline!FEATURE_FIND_XDEV! }
    \label{fig:screenshowt_partial}
\end{subfigure}
\hfill
\begin{subfigure}[t]{0.3\textwidth}
    \includegraphics[width=\textwidth]{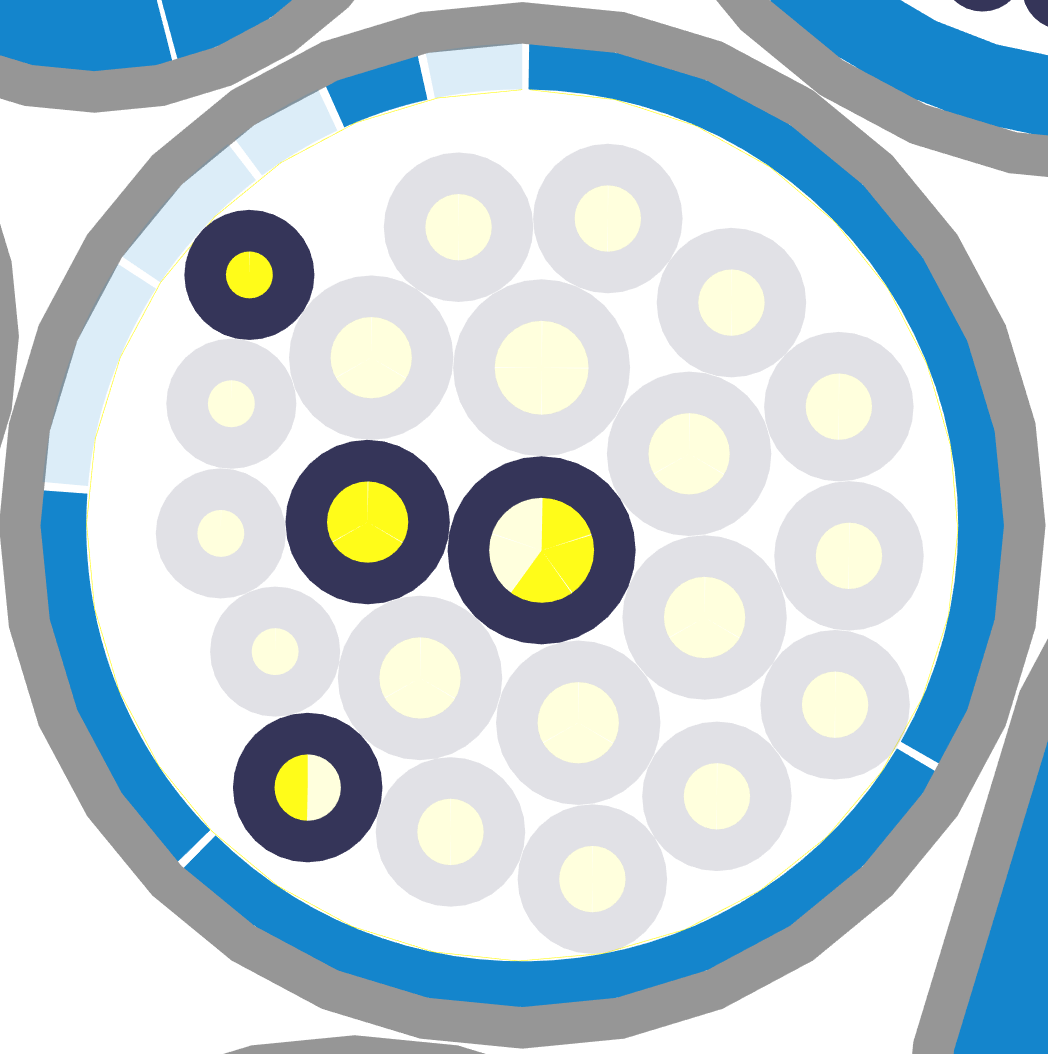}
    \subcaption { All \lstinline!find! specific features are disabled }
    \label{fig:screenshowt_empty}
\end{subfigure}
\caption{Visualizing the structure of BusyBox's ``find.c'' with three different configurations}
\label{fig:conditional_compilation}
\end{figure*}

Figure~\ref{fig:conditional_compilation} shows the structure of the translation unit \lstinline!find.c! with three different configurations.
Our tool improves the developers understanding of the resulting structur and behavior by making the commonalities and differences explicit.

\section{Discussion}

With our prototype we focus on the visualization design to support developers when exploring software systems which are making use of conditional compilation. 
We have therefore placed more emphasis on usability than on feature completeness.
The application scenario demonstrates how the visualization supports the developer's usual workflow and eliminates time-consuming steps.
Q2 can already be answered completely.
Q1 can only be answered partially since not all feature locations are visually detectable.
The necessary information is already available, but is not yet accessible in the user interface.
For example, when selecting a feature, it would be possible to highlight all methods affected by the feature.
It would also be helpful to support saving and loading configurations for subsequent analyses as well as comparing complete configurations visually.
As soon as these features are implemented, we will compare our tool with existing solutions.

As for many software visualizations scalability is a critical point and necessary to use the tool in practice.
The generation process took one day on a conventional notebook.
We can visualize systems with up to four million lines of code without any problems. 
If the visualization becomes too complex, performance may decrease.
However, the visualization framework still offers a lot of potential to improve performance to visualize larger systems.

\section{Conclusion}

Our tool supports the developer to explore variability implemented with CPP, especially in the context of SPLs.
It simplifices tracing and understanding the effect of enabling or disabling these flags with respect to the code compiled subsequently.
Thus, it bridges the gap between feature models and diagrams as more abstract representations of variability and its concrete implementation with the means of CPP. 
However, some features are still missing for use in practice that need to be addressed in future work.


%
\bibliographystyle{ACM-Reference-Format}
\bibliography{literature,mendeley}

\end{document}